\documentclass[12pt]{article}

\usepackage{times}

\usepackage{graphicx}

\newenvironment{sciabstract}{%
\begin{quote} \bf}
{\end{quote}}

\newcounter{lastnote}

\title{Comment on "Observation of the Wigner-Huntington transition to metallic hydrogen''}

\begin{document} 
\baselineskip24pt


\maketitle

\author{Xiao-Di Liu$^1$, Philip Dalladay-Simpson$^2$, Ross T. Howie$^2$, Bing Li$^{2,3}$ and Eugene Gregoryanz$^{1,2,4}$} 
\\

\normalsize{$^1$Key Laboratory of Materials Physics, Institute of Solid State Physics,  CAS, Hefei, China}
\\
\normalsize{$^2$Center for High Pressure Science and Technology Advanced Research, Shanghai,  China}

\normalsize{$^3$Center for the Study of Matter at Extreme Conditions, Department of Mechanical and Materials Engineering, 
Florida International University, Miami, USA}
\\
\normalsize{ $^4$Centre for Science at Extreme Conditions and School of Physics and Astronomy, University of Edinburgh, Edinburgh, U.K.}\\

\date{}


\baselineskip24pt


\maketitle


\begin{sciabstract}
Dias and Silvera  (Letters, p. 715, 2017)  claim the observation of the Wigner-Huntington transition to metallic hydrogen at 
495 GPa. We show that neither the claims of the record pressure or the phase transition to a metallic state are supported by any data 
and contradict the authors' own unconfirmed previous results. 
\end{sciabstract}

Ref. (1)  presents only one experimental run claiming the record pressure of 495 GPa. The paper presents 
2 figures of the phase diagram, iPhone photos of the sample, and the deduced reflectivity of 4 wavelengths only at the highest pressure point. The supplementary 
materials provide the processed and fitted infra-red (IR) absorption spectra from 135 to 335 GPa,  the pressure versus force curve assuming the linear dependence, and the Raman 
spectra of the stressed diamond from 1200 to 2200 cm$^{-1}$. The absence of data combined with the uncritical claims have led to the unprecedented 3 Comments 
(2-4) written within a month of publication of Ref. (1).

In the past 5 years we have conducted  $\sim$120 experiments on hydrogen reaching above 200 GPa (5-10).
In $\sim$30 runs out of 120 the pressure exceeded 300 GPa and in only 5 out of 120 the pressure exceeded 350 GPa. The extensive statistics show that the diamond 
culet sizes of 30 $\mu$m diameter (used in Ref. (1)) could be used to reach maximum pressures of $\sim$315$\pm$10 GPa with the probability of 20$\%$.  In order to 
reach pressures close to 400 GPa, with lower probability of 10$\%$, the culet sizes of 15 $\mu$m must be used with the sample contracting to 2-3 $\mu$m at the 
highest pressure. Our statistics are in an excellent agreement with other groups working on hydrogen at high pressure (11-13). The authors of Ref. (1) had 2 experimental runs 
in the past year using culets of 30 $\mu$m diameter (14,1) claiming the unsubstantiated pressures of  420 and 495 GPa.

Ref. (1) claims that a combination of annealing, fine polishing and coating the culet with Al$_2$O$_3$ has led to the increase in maximum pressure with larger 
culets (and samples) compared with previous studies (5-13). Those techniques would likely decrease the probability of the premature 
failure due to hydrogen diffusion, but there is no supportive evidence that these techniques would improve the mechanical stability of diamond.
The record pressure of 600 GPa was recently shown on metallic samples {\it i.e.} Re and/or Pt  using
novel double stage approach with the pressure generating area having the linear size of 3 $\mu$m only (15) (100 times smaller working area than in (1)).
        
Fig. 1 shows the pressure versus load curve from (1) and our own data. Ref. (1) uses 3 {\it different, non-overlapping calibration methods}
at $\sim$100 and $\sim$300 GPa. Measuring pressure by estimating the load is not a direct method, as it does not probe the sample 
or/and calibrant {\it in situ}. The loading curve is {\it unique} for each experimental run and depends {\it only} on the sizes of the culets, angles of 
the bevels, compressibility of the gasket and sample. The dependency cannot be linear as shown in (1) but always 
consists of 3 distinct regimes, each of which is sub-linear with differing gradients: plastic deformation of the gasket,  fast rise of pressure beyond 
the plastic deformation, followed by the much slower pressure increase (saturation) due to the bending ("cupping") of the diamonds. 

In order to {\bf make} the dependence linear, Ref. (1) rescales the point from 420 to 400 GPa taken from a {\it different experiment} (14) and uses  
4 meaningless points of "visual observations".  The last point of 495 GPa (as well as 420 GPa point from (14))  is deduced from the single Raman spectrum 
which does not cover the wide energy range showing the signal from hydrogen and/or pressure induced fluorescence. This rules out any critical 
assessment that the sample did not diffuse out and that the assigned peak of the stressed diamond is not due to any other factors. 

The IR absorption data are consistent with the loss of the sample and contradict the authors' own previous claims (14). In Fig. 1 we 
plot the raw data provided by but not plotted in (1) together with the normalised spectra. The sample was clearly diffusing out between 
the claimed pressures of 314 and 338 GPa and completely lost above this.  The lack of IR transmission above 338 GPa and that the "sample" 
appears dark at 415 GPa correlates with the gasket flow scenario. In (14) the authors present the same IR data up to 420 GPa, but do not provide 
any explanation why their results are so different in two consecutive experiments. Also, no explanation is given as to why there was no IR 
transmission/absorption between 335 and 495 GPa, the pressure range in which hydrogen is semiconducting.

The analysis of the photos provided is also very consistent with the loss of the sample at an earlier stage in the experiment. The sample at 415 GPa is 
expanded under pressure and is slightly {\it bigger} than at 205 GPa, while it is the same size at 495 GPa. In (8) we demonstrate the photos of the 
sample up to 280 GPa which show the flow of metal in the sample chamber and diffusion of H$_2$ out of the chamber. As the chamber closes, 
the thin layer of hydrogen between the collapsed gasket and the diamond gives the dark appearance (compare Fig. 2B in (1) and  Fig. 3 in (8)). 
With the increasing pressure all hydrogen diffuses out of the chamber and the dark area becomes shiny as it bridges the anvils but would have 
different reflectivity from the rest of the gasket as in Fig. 2C in (1) (see also photo of "{\it metallic hydrogen}" in Fig. 3D in (11)).    
 
The presented phase diagram (Figs. 1 and 4) show an unconfirmed {\it H$_2$-PRE phase} for which no data are presented in the paper. The existence of this phase 
was surmised in (14) and even though no supporting data are given, Ref. (1) states "{\it after the 335 GPa pressure point resulted in our sample 
starting  to turn {\bf black} (Fig. 2B) as it transitioned into the H2-PRE phase}" which the authors claimed to be transparent (see Fig. 3S in (14)). 
Furthermore, Ref. (1) cites one of the authors' own publication on melting  but, without any explanation, plots the smooth melting curve (see Fig. 1) 
contradictory to the authors' cited paper which claims the existence of the sharp peak at 60 GPa.

\begin{quote}
{\bf References and Notes}

\begin{enumerate}

\item R. Dias and I. Silvera, {\it Science} {\bf 355},  715 (2017).
\item A. Goncharov and V. Struzhkin, arXiv:1702.04246 (2017)
\item M. Eremets and I. Frozdov, arXiv:1702.05125 (2017)
\item P. Loubeyere {\it et al}., arXiv:1702.07192 (2017)
\item R. Howie {\it et al}., {\it Phys. Rev. Lett.},  {\bf 108}, 125501 (2012)
\item R. Howie {\it et al}., {\it Phys. Rev. B},  {\bf 86}, 214104  (2012)
\item R. Howie {\it et al}., {\it J. Appl. Phys},   {\bf 114 },  073505 (2013)
\item R. Howie {\it et al}., {\it Phys. Rev. Lett.},  {\bf  113}, 175501 (2014)
\item R. Howie {\it et al}., {\it Nat. Mat},  {\bf 14}, 495 (2015)
\item P. Dalladay-Simpson {\it et al.}  {\it Nature}, {\bf  529}, 63 (2016)
\item M. Eremets and I. Troyan {\it Nat. Mat.}, {\bf 10}, 927 (2011)
\item P. Loubeyre {\it et al., Phys. Rev. B},  {\bf 87}, 134101 (2013)
\item C. Zha {\it et al., Phys. Rev. Lett.}, {\bf 110}, 217402 (2013)
\item R. Dias  {\it et al}., 	arXiv:1603.02162  (2016)
\item  L. Dubrovinsky {\it Nat. Comm.} {\bf 3}, 1163 (2012)
\end{enumerate}
\end{quote}

\newpage

\begin{figure}
\begin{center}
\includegraphics[scale=0.5]{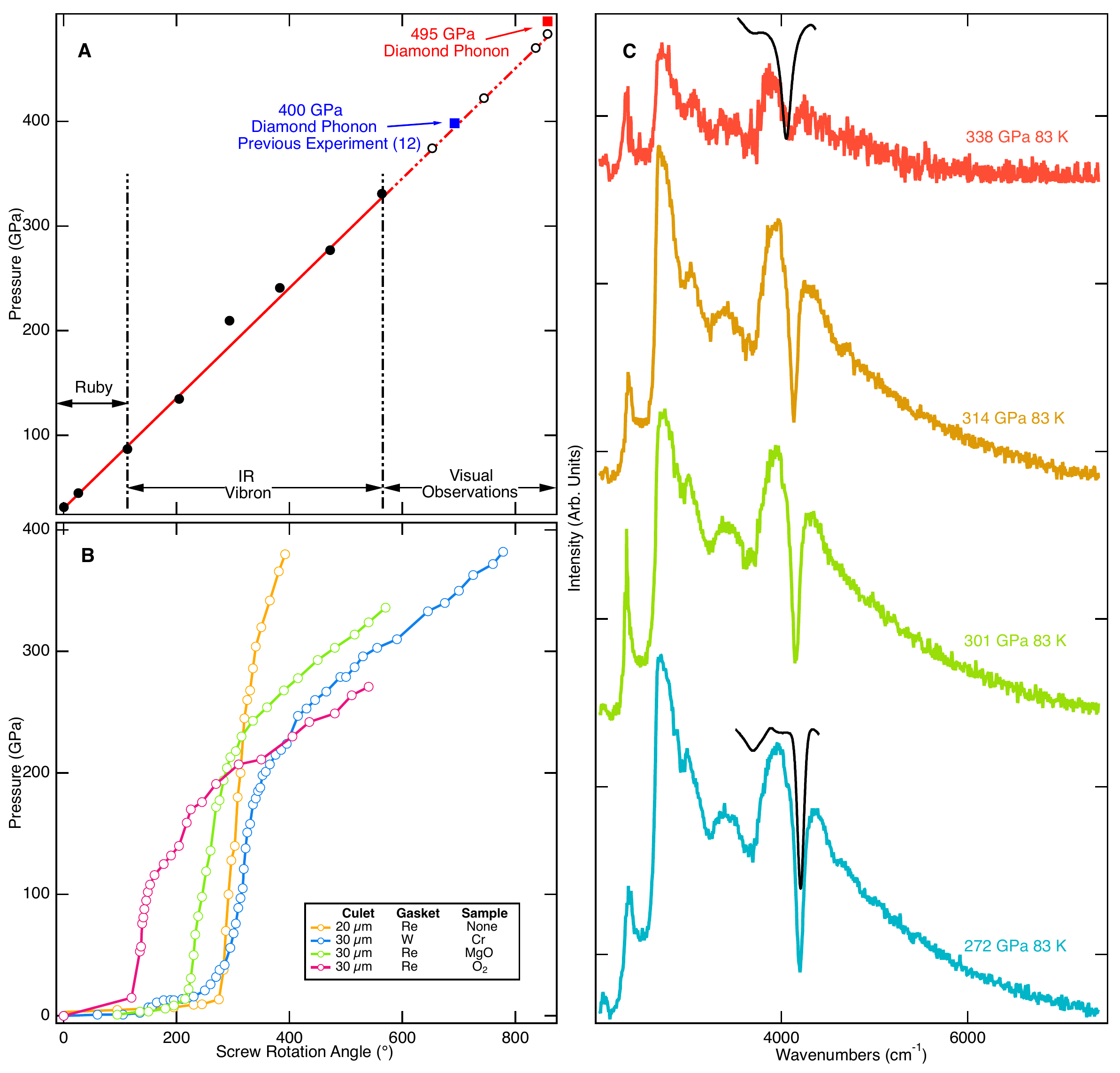}
\caption{Pressure versus load: (A) Ref. (1), (B) our data. (C) Some raw IR absorption spectra provided but not plotted in Ref. (1) (in colour). The black curves 
are normalised spectra taken from Fig. S1 in (1). The normalised spectrum at 335 GPa looks artificially "enhanced" and shifted to higher pressures by the authors.}
\label{Fig1}
\end{center}
\end{figure}

\begin{figure}
\begin{center}
\includegraphics[scale=0.55]{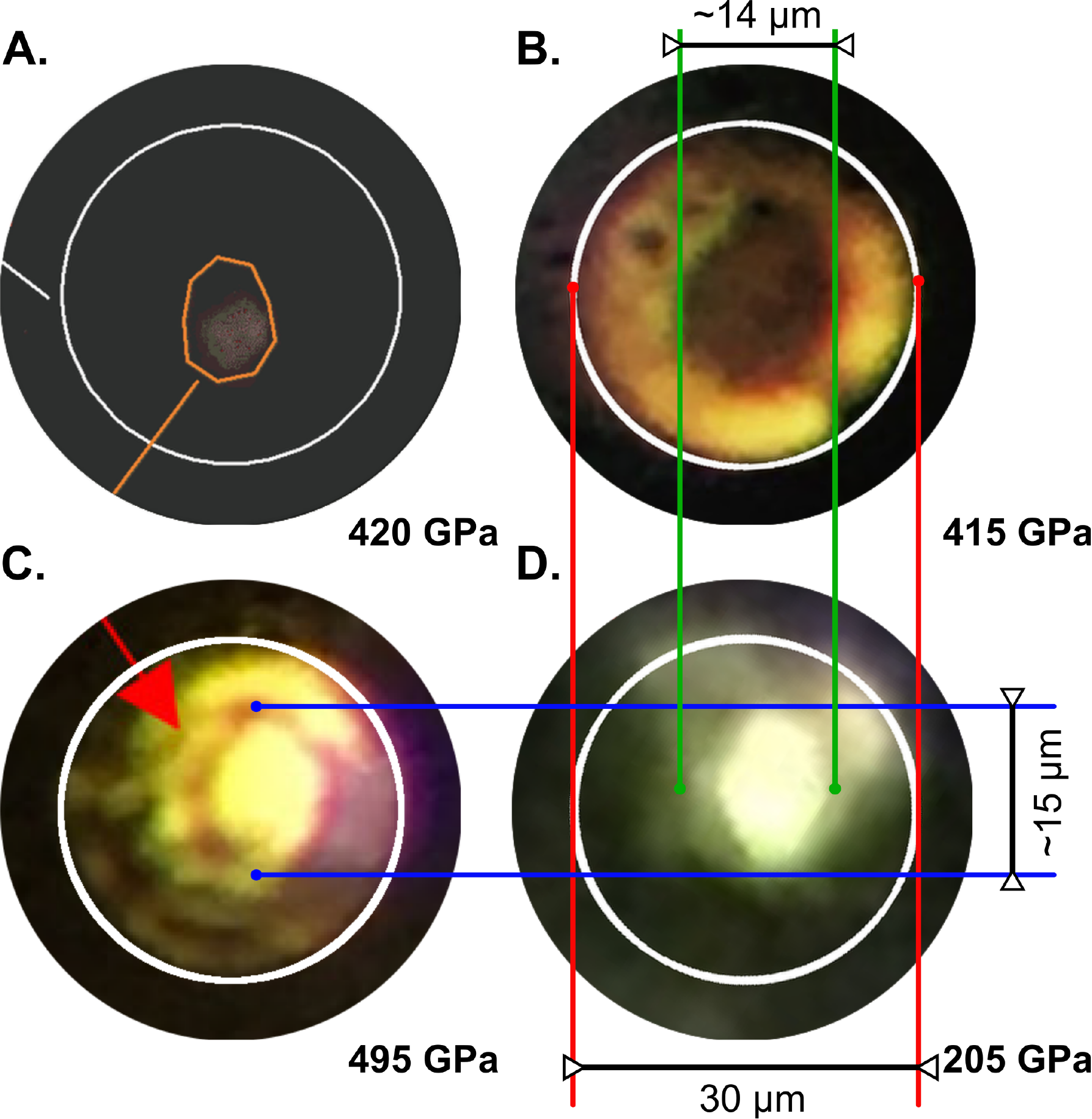}
\caption{ A. Photo of the semitransparent sample at 420 GPa from (14). B. Photo of the dark sample at 415 GPa from (1).
C and D. Photos of the samples at 205 and 495 GPa from (1). We have used the 30 $\mu$m stated by (1) being the size of 
the culet as the scale to estimate the size of the sample. In the vertical direction the sample reaches 15 $\mu$m which is 
almost twice larger than than the 8-10 $\mu$m stated in (1). For comparison, the size of the hydrogen sample at 390 GPa was $\sim$2$\mu$m;
see Fig. 3 in (10).}
\label{Fig2}
\end{center}
\end{figure}















\end{document}